\documentclass[12pt,a4paper]{article}
\usepackage[OT2,T1]{fontenc}
\usepackage{amssymb}
\usepackage{amsmath}
\usepackage{stmaryrd}
\usepackage{mathrsfs}
\usepackage{mathtools}
\usepackage{bm}
\usepackage{graphicx}
\usepackage{here}
\usepackage[margin=1in]{geometry}
\usepackage{cite}
\usepackage{sectsty}
\allsectionsfont{\sffamily}
\DeclareMathAlphabet{\mathscrbf}{OMS}{mdugm}{b}{n}

\let\OLDthebibliography\thebibliography
\renewcommand\thebibliography[1]{
  \OLDthebibliography{#1}
  \setlength{\parskip}{0pt}
  \setlength{\itemsep}{6pt plus 0.3ex}
}

\usepackage{tikz}
\usetikzlibrary{arrows,calc}
\newcommand{\batman}{\begin{tikzpicture}[scale=0.045]
    \draw[fill,black]   (-0.25,1.48) .. controls (-0.1,1.6) and (0.1,1.6) .. (0.25,1.48) -- (0.42,1.88) .. 
controls (0.425,1.8) and (0.41,1.3) .. (0.45,1.2) .. controls (0.6,1.05) and (1.96,1.05) .. (1.98,2.08) -- 
(5.93,2.08) .. controls (4.2,1.45) and (4,0.3) .. (4.2,-0.28) .. controls (2.4,-0.09) and (0.4,-0.5) .. 
(0,-2.052) .. controls (-0.4,-0.5) and (-2.4,-0.09) .. (-4.2,-0.28) .. controls (-4,0.3) and (-4.2,1.45) .. 
(-5.93,2.08) -- (-1.98,2.08) .. controls (-1.96,1.05) and (-0.6,1.05) .. (-0.45,1.2) .. controls (-0.41,1.3) 
and (-0.425,1.8) .. (-0.42,1.88) -- (-0.25,1.48);
\end{tikzpicture}}

\usepackage{tikz}
\usetikzlibrary{arrows,calc}
\newcommand{\shadowbatman}{\begin{tikzpicture}[scale=0.045]
    \draw[fill,white]   (-0.25,1.48) .. controls (-0.1,1.6) and (0.1,1.6) .. (0.25,1.48) -- (0.42,1.88) .. 
controls (0.425,1.8) and (0.41,1.3) .. (0.45,1.2) .. controls (0.6,1.05) and (1.96,1.05) .. (1.98,2.08) -- 
(5.93,2.08) .. controls (4.2,1.45) and (4,0.3) .. (4.2,-0.28) .. controls (2.4,-0.09) and (0.4,-0.5) .. 
(0,-2.052) .. controls (-0.4,-0.5) and (-2.4,-0.09) .. (-4.2,-0.28) .. controls (-4,0.3) and (-4.2,1.45) .. 
(-5.93,2.08) -- (-1.98,2.08) .. controls (-1.96,1.05) and (-0.6,1.05) .. (-0.45,1.2) .. controls (-0.41,1.3) 
and (-0.425,1.8) .. (-0.42,1.88) -- (-0.25,1.48);
\end{tikzpicture}}

\usepackage[colorlinks,allcolors=blue]{hyperref}

\newcommand{\be}{\begin{equation}}
\newcommand{\ee}{\end{equation}}
\newcommand{\ben}{\begin{enumerate}}
\newcommand{\een}{\end{enumerate}}
\newcommand{\bi}{\begin{itemize}}
\newcommand{\ei}{\end{itemize}}
\newcommand{\bmm}{\begin{pmatrix}}
\newcommand{\emm}{\end{pmatrix}}

\newcommand{\Ad}{\text{Ad}}
\newcommand{\bms}{\mathfrak{bms}_3}
\newcommand{\bra}{\langle}

\newcommand{\demi}{\frac{1}{2}}
\newcommand{\der}{\partial}
\newcommand{\Diff}{\text{Diff}\,S^1}

\newcommand{\eg}{e.g.\ }

\newcommand{\hDiff}{\widehat{\text{Diff}}\,S^1}

\newcommand{\hVect}{\widehat{\text{Vect}}\,S^1}
\newcommand{\ie}{i.e.\ }
\newcommand{\ket}{\rangle}
\newcommand{\nn}{\nonumber}
\newcommand{\phii}{\varphi}

\renewcommand{\refeq}[1]{\stackrel{\text{(\ref{#1})}}{=}}

\newcommand{\Vect}{\text{Vect}\,S^1}

\newcommand{\cO}{{\cal O}}

\newcommand{\cS}{{\cal S}}

\newcommand{\cU}{\,{\cal U}}

\newcommand{\mg}{\mathfrak{g}}
\newcommand{\cs}{\mathfrak{s}}
\newcommand{\cu}{\mathfrak{u}}

\newcommand{\sI}{\mathscr{I}}

\newcommand{\CC}{\mathbb{C}}

\newcommand{\NN}{\mathbb{N}}
\newcommand{\RR}{\mathbb{R}}
\newcommand{\ZZ}{\mathbb{Z}}

\begin{document}

\hrule
\begin{center}
\Large{\bfseries{\textsf{Thomas Precession for Dressed Particles}}}
\end{center}
\hrule
~\\

\begin{center}
\large{\textsf{$^{\shadowbatman}$Blagoje Oblak$^{\batman}$}}
\end{center}
~\\

\begin{center}
\begin{minipage}{.9\textwidth}\small\it
\begin{center}
Institut f\"ur Theoretische Physik\\
ETH Z\"urich\\
CH-8093 Z\"urich, Switzerland
\end{center}
\end{minipage}
\end{center}

\vspace{1cm}

\begin{center}
\begin{minipage}{.9\textwidth}
\begin{center}{\bfseries{\textsf{Abstract}}}\end{center}
We consider a particle dressed with boundary gravitons in three-dimensional Minkowski space. The existence of BMS transformations implies that the particle's wavefunction picks up a Berry phase when subjected to changes of reference frames that trace a closed path in the asymptotic symmetry group. We evaluate this phase and show that, for BMS superrotations, it provides a gravitational generalization of Thomas precession. In principle, such phases are observable signatures of asymptotic symmetries.
\end{minipage}
\end{center}

\vfill
\noindent
\mbox{}
\raisebox{-3\baselineskip}{%
\parbox{\textwidth}{\mbox{}\hrulefill\\[-4pt]}}
{\scriptsize$^{\batman}$ E-mail: boblak@phys.ethz.ch}

\thispagestyle{empty}
\newpage

\textsf{\tableofcontents}
~\\
\hrule


\section{Introduction}

Consider a spinning particle that follows an accelerated trajectory --- for instance an electron revolving around an atomic nucleus. Seen from the particle's locally inertial reference frame, acceleration is interpreted as a continuous sequence of infinitesimal Lorentz boosts. If these boosts are not collinear, they change both the particle's velocity and the orientation of its reference frame, resulting in a rotation that affects its spin vector; for periodic motion this leads to the phenomenon of Thomas precession \cite{Thomas:1926dy} that is routinely observed in atomic spectra \cite{Uhlenbeck}. Abstractly, the effect is due to Wigner rotations \cite{Wigner:1939cj} that appear in representations of the Poincar\'e group, so it is purely kinematical --- it follows solely from space-time symmetries. The purpose of this paper is to investigate the generalization of Thomas precession that emerges when Poincar\'e is replaced by the Bondi-van der Burg-Metzner-Sachs (BMS) group \cite{Bondi:1962px}.\\

The BMS group is an infinite-dimensional extension of Poincar\'e describing the asymptotic symmetries of gravitational systems that become Minkowskian at (null) infinity.\footnote{It was recently shown that BMS symmetry can also be defined at spatial infinity \cite{Troessaert:2017jcm}.} Just as Poincar\'e comprises Lorentz transformations and translations, BMS consists of superrotations \cite{Barnich:2009se} and supertranslations. A similar symmetry enhancement occurs in three-dimensional Anti-de Sitter (AdS$_3$) gravity, whose asymptotic symmetries span two copies of the Virasoro group that extend SL$(2,\RR)$ isometries \cite{Brown:1986nw}. Owing to the relation between asymptotic symmetries and soft theorems \cite{Strominger:2013jfa}, irreducible unitary representations of BMS may be regarded as particles dressed with soft/boundary gravitons. In four dimensions and without superrotations, these particles were classified long ago in \cite{McCarthy01}, while for three-dimensional BMS (with superrotations) \cite{Ashtekar:1996cd} they were obtained in \cite{Barnich:2014kra}.\\

In this work we describe the asymptotic extension of Thomas precession for dressed particles in three space-time dimensions. In that setting, BMS superrotations have a smooth action on null infinity (as opposed to their singular behaviour in four dimensions \cite{Barnich:2011mi}) and the entire group structure is mathematically well-defined. In fact, as in the AdS$_3$ case of \cite{Oblak:2017ect}, we will derive a general formula for Berry phases \cite{Berry:1984jv} associated with closed curves in the space of {\it all} asymptotic symmetries (including supertranslations). The computation will rely on certain technical tools developed in \cite{Oblak:2017oge} and will reproduce the `geometric action functionals' of BMS$_3$ \cite{Barnich:2017jgw}. For superrotations this will provide a gravitational, spin-dependent extension of Thomas precession (eq.\ (\ref{s23}) below), while for supertranslations it will yield a Hannay angle \cite{Hannay} involving the Planck mass (eq.\ (\ref{ss19}) below). In principle, such and similar effects may be observable either in actual gravitational experiments (where they are related to gravitational memory \cite{zel1974radiation,Strominger:2014pwa}), or in two-dimensional, possibly non-relativistic \cite{Bagchi:2009pe}, conformal field theories (CFTs).\\

The plan is as follows. Section \ref{sec2} is a brief review of BMS symmetry in three dimensions and of the corresponding description of dressed particles. This is used in section \ref{sec3} to derive the Berry phase picked by a particle subjected to changes of reference frames that trace a loop in the BMS$_3$ group. We conclude in section \ref{sec4} with a discussion of plausible applications and extensions of our results, including their four-dimensional version.

\section{BMS$_3$ particles}
\label{sec2}

In this section we recall the definition of the BMS$_3$ group \cite{Ashtekar:1996cd}, including central extensions \cite{Barnich:2006av,Barnich:2011ct}, and the ensuing classification of irreducible unitary representations \cite{Barnich:2014kra}. Just as relativistic particles, BMS$_3$ particles are labelled by their mass and spin, but their larger Hilbert space accounts for gravitational dressing. We refer to \cite{Oblak:2016eij} for a much more thorough exposition of these matters.

\subsection{BMS$_3$ as an extension of Poincar\'e}

\begin{figure}[t]
\centering
\includegraphics[width=0.30\textwidth]{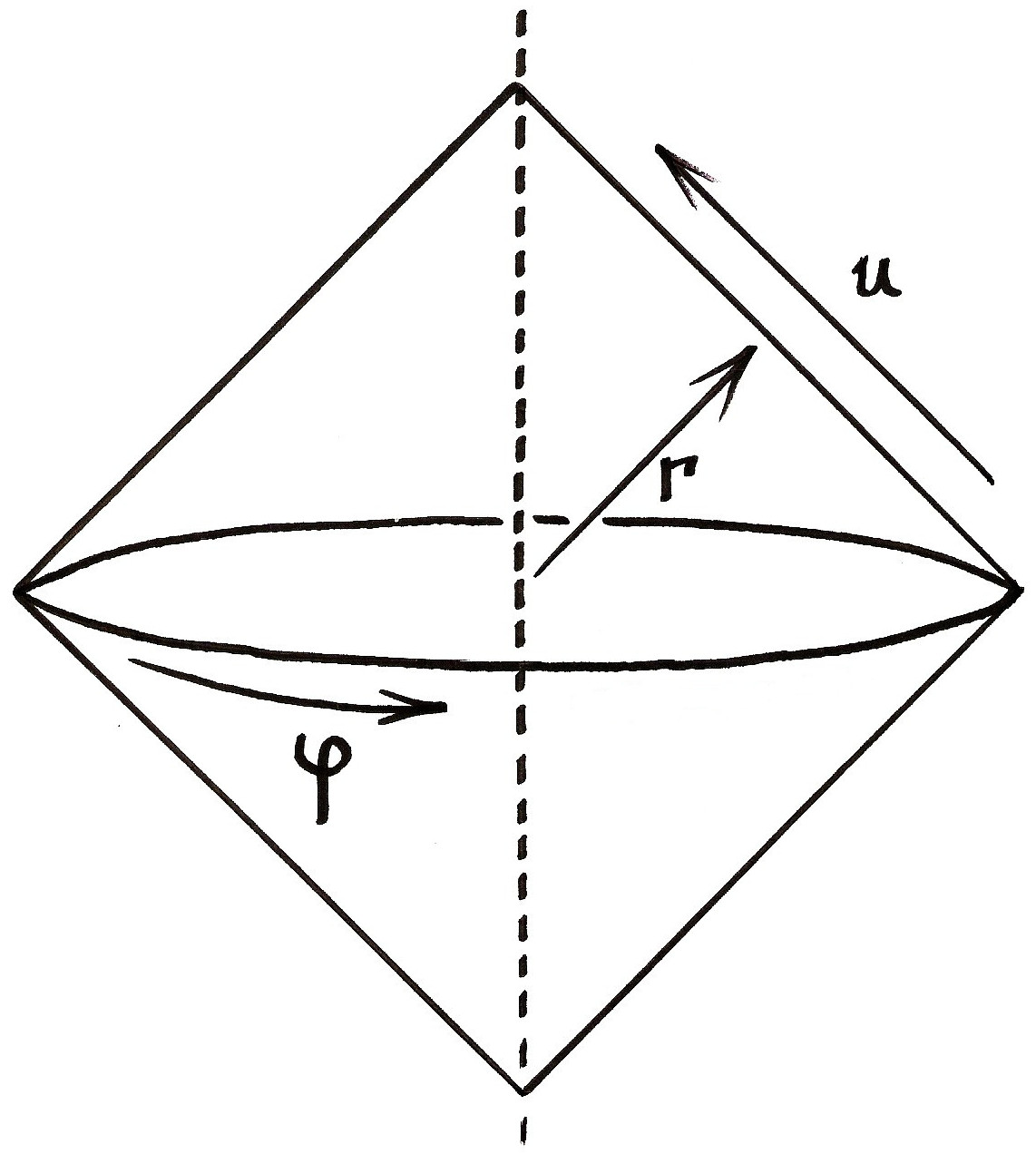}
\caption{Bondi coordinates in Minkowski space; future null infinity ($\sI^+$) is the region $r\rightarrow\infty$ spanned by $(u,\phii)$. Poincar\'e and BMS transformations act on $\sI^+$ by diffeomorphisms (\ref{t6}) that stretch celestial circles and deform slices of constant retarded time.\label{s6}}
\end{figure}

To motivate the definition of BMS$_3$ transformations, consider three-dimensional Minkowski space-time. It is a manifold $\RR^3$ whose metric in inertial coordinates reads $\text{d}s^2=\eta_{\mu\nu}\text{d}x^{\mu}\text{d}x^{\nu}$, where $\mu,\nu=0,1,2$ and $\eta_{\mu\nu}=\text{diag}(-1,1,1)$. In terms of retarded Bondi coordinates $(r,u,\phii)$ given by
\be
r
=
\big[(x^1)^2+(x^2)^2\big]^{1/2},
\qquad
u
=
x^0-r,
\qquad
e^{i\phii}
=
\frac{x^1+ix^2}{r},
\label{tt6}
\ee
future null infinity $\sI^+$ is the region $r\rightarrow\infty$ with finite retarded time $u$; see fig.\ \ref{s6}. Poincar\'e transformations are isometries of Minkowski space-time, and they admit a well-defined large $r$ limit. Indeed one can show that they act on $\sI^+$ as diffeomorphisms
\be
\phii\mapsto f(\phii),
\qquad
u\mapsto f'(\phii)u+\alpha(f(\phii))
\label{t6}
\ee
where $\alpha(\phii)=\alpha^0-\alpha^1\cos\phii-\alpha^2\sin\phii$ is a space-time translation, while the Lorentz transformation $f$ is specified by complex coefficients $\alpha,\beta$ such that
\be
e^{if(\phii)}
=
\frac{\alpha e^{i\phii}+\beta}{\bar\beta e^{i\phii}+\bar\alpha},
\qquad |\alpha|^2-|\beta|^2=1.
\label{ss6}
\ee
In this language, rotations read $f(\phii)=\phii+\theta$ while Lorentz boosts towards $\phii=0$ are given by $\alpha=\cosh(\lambda/2)$, $\beta=\sinh(\lambda/2)$ in terms of the rapidity $\lambda$.\\

When dealing with gravity, one prescribes fall-off conditions for the metric at null infinity. The asymptotic symmetry group then consists of space-time diffeomorphisms that preserve these fall-offs. For those of \cite{Ashtekar:1996cd} this yields the BMS$_3$ group \cite{Barnich:2014kra},\footnote{See \cite{Detournay:2016sfv} for alternative fall-offs; the resulting asymptotic symmetries always contain (\ref{s7}).}
\be
\begin{split}
\text{BMS}_3
\equiv
\Diff\ltimes\Vect.\quad\quad\quad\quad\quad\\[-.1cm]
\swarrow\quad\quad\quad\quad\quad\searrow\quad\quad\quad\quad\quad~\\[-.1cm]
\;\;\;\,\quad\quad\qquad\text{superrotations}\quad\quad\quad\text{supertranslations}
\end{split}
\label{s7}
\ee
Its elements are pairs $(f,\alpha)$ acting on $\sI^+$ as in (\ref{t6}), but with fewer restrictions: one only assumes that $\alpha(\phii)$ is $2\pi$-periodic and that $f(\phii)$ is an orientation-preserving diffeomorphism, so $f'(\phii)>0$ and $f(\phii+2\pi)=f(\phii)+2\pi$. The group operation is that of a semi-direct product,
\be
(f,\alpha)\cdot(g,\beta)
=
(f\circ g,\alpha+f\cdot\beta),
\label{ss8}
\ee
where $f\cdot\beta$ is the action of a diffeomorphism on a vector field on the circle.\footnote{This action is actually the adjoint representation of $\Diff$, so $f\cdot\beta\equiv\Ad_f\beta$.} For brevity we will sometimes write $f\circ g\equiv fg$, omitting the composition. For quantum-mechanical applications one needs to take central extensions into account; in the case at hand, the universal central extension of (\ref{s7}) is
\be
\widehat{\text{BMS}}{}_3
=
\hDiff\ltimes\hVect
\label{ss7}
\ee
where $\hDiff$ is the Virasoro group and $\hVect$ is its Lie algebra, seen as an Abelian vector group. The group operation is a suitable extension of (\ref{ss8}) --- see eq.\ (\ref{s18}).\\

The two hats on the right-hand side of (\ref{ss7}) mean that BMS$_3$ symmetry admits two central charges $c_1$, $c_2$. Indeed, denoting superrotation and supertranslation generators by $J_m$ and $P_m$ respectively ($m\in\ZZ$), it is customary to write their commutation relations as
\begin{align}
[J_m,J_n] &= (m-n)J_{m+n}+\frac{c_1}{12}m(m^2-1)\delta_{m+n,0}\nn\\
\label{t7}
[J_m,P_n] &= (m-n)P_{m+n}+\frac{c_2}{12}m(m^2-1)\delta_{m+n,0}\\
[P_m,P_n] &= 0.\nn
\end{align}
These expressions coincide with the Poisson brackets of surface charges associated with BMS$_3$ transformations. The dimensionless central charge $c_1$ vanishes in Einstein gravity, but it is non-zero in parity-breaking theories such as topologically massive gravity \cite{Deser:1981wh}. As for $c_2$, it is best seen as a dimensionful mass scale; with the normalization of \cite{Barnich:2006av} it is essentially the Planck mass in three dimensions,
\be
c_2=3/G
\label{s8}
\ee
where $G$ is Newton's constant. From now on we always set $c_2>0$, while $c_1$ is arbitrary.

\subsection{BMS$_3$ particles}

Taking BMS seriously as a symmetry group for particle physics, it is natural to investigate its irreducible unitary representations.\footnote{In three dimensions, BMS evades the Coleman-Mandula theorem \cite{Coleman:1967ad} thanks to the fact that each BMS$_3$ particle contains an infinity of Poincar\'e representations with increasing mass.} In four dimensions, and with superrotations restricted to their Lorentz subgroup, this question was addressed in \cite{McCarthy01} and its follow-ups. In three dimensions the unitary representations of (\ref{ss7}) were described in \cite{Barnich:2014kra};\footnote{See also \cite{Bagchi:2009pe} for early considerations from the Galilean perspective.} we refer to them as {\it BMS$_3$ particles}. Thomas precession will be a Berry phase that affects the wavefunctions of such particles, so we briefly review their properties here.\\

Owing to the semi-direct product structure of (\ref{s7})-(\ref{ss7}), the classification of BMS$_3$ particles is similar to that of Poincar\'e representations in terms of mass and spin \cite{Wigner:1939cj}. The subtlety is to understand what one means by `mass'. Indeed, the mass of a relativistic particle is specified by its energy-momentum $p_{\mu}=(E,\textbf{p})$, which belongs to the dual of the vector group of space-time translations. But for BMS$_3$, translations are enhanced to {\it supertranslations}, which are arbitrary functions $\alpha(\phii)$ on the celestial circle. Their duals are {\it supermomenta}, each of which is a function $p(\phii)$ paired with $\alpha$'s according to
\be
\langle p,\alpha\rangle
=
\frac{1}{2\pi}\oint\text{d}\phii\,p(\phii)\alpha(\phii)
\label{s9}
\ee
where the integral over $\phii$ runs from $0$ to $2\pi$. Since constant $\alpha$'s generate time translations, the energy of $p(\phii)$ is its zero-mode; similarly, the lowest two Fourier modes of $p(\phii)$ are components of spatial momentum. Higher modes are genuine gravitational contributions that do not appear in the Poincar\'e group. From a general-relativistic perspective, supermomentum coincides with the Bondi mass aspect of asymptotically flat metrics \cite{Barnich:2015uva}.\\

A key ingredient in the classification of Poincar\'e representations is the notion of `mass shells' or momentum orbits. Each orbit $\cO_p$ consists of an energy-momentum vector $p$, together with all vectors $q$ that are obtained by acting on $p$ with Lorentz transformations. In the same way, the classification of BMS$_3$ particles relies on orbits of supermomenta under superrotations. Each orbit is built by fixing a supermomentum function $p(\phii)$ and a value for the dimensionful central charge $c_2$, then acting on $p$ with superrotations. This action coincides with the transformation law of (chiral) CFT stress tensors under conformal transformations \cite{Barnich:2014kra}:
\be
\big(f\cdot p\big)(f(\phii))
=
\frac{1}{(f'(\phii))^2}\Big[p(\phii)+\frac{c_2}{12}\,S[f](\phii)\Big],
\label{s10}
\ee
where $S[f]=f'''/f'-\tfrac{3}{2}(f''/f')^2$ is the Schwarzian derivative of $f$. Abstractly, eq.\ (\ref{s10}) is the coadjoint representation of the Virasoro group, whose orbits were classified in \cite{Lazutkin} (see also \cite{Witten:1987ty}); they are labelled by a parameter $M$ defined through the Wilson loop
\be
\text{Tr}\Bigg(P\exp\bigg[\oint\text{d}\phii\bmm 0 & 1 \\ 6p(\phii)/c_2 & 0 \emm\bigg]\Bigg)
=
2\cos\bigg(\pi\sqrt{1-\frac{24M}{c_2}}\,\bigg).
\label{ss10}
\ee
This relation determines the mass $M$ of a dressed particle from the knowledge of its supermomentum $p(\phii)$; it is invariant under superrotations, as the same value of $M$ is given by $p$ and $f\cdot p$ for any $f\in\Diff$. It is a BMS$_3$ analogue of the Poincar\'e-invariant definition $M=\sqrt{-p_{\mu}p^{\mu}}$. Note that the central charge $c_2$ fixes a mass scale; indeed the conjugacy class of the path-ordered exponential in (\ref{ss10}) changes from elliptic to hyperbolic when $M$ crosses the critical value $M=c_2/24$. In gravity this corresponds to a bifurcation between conical deficits ($0<M<c_2/24$) and flat space cosmologies ($M>c_2/24$) \cite{Cornalba:2002fi}.\\

In this paper we focus on massive BMS$_3$ particles ($M>0$), though our results will also hold for the vacuum representation that has $M=0$ and is based on the orbit of $p(\phii)=-c_2/24$. (We will not consider tachyons or massless particles.) For $M>0$, the particle admits a rest frame and its little group U$(1)$ consists of pure rotations $f(\phii)=\phii+\theta$; it is the subgroup of superrotations that preserves the rest frame. Following Wigner's classification of Poincar\'e representations \cite{Wigner:1939cj}, this means that the spin of any massive particle is a real number $s\in\RR$ that specifies a (generally projective) unitary, irreducible representation of U$(1)$. In fact, more appropriately, the central extension of superrotations implies that the little group also contains shifts along the direction dual to the central charge $c_1$. As a result, the little group representation is really labelled by {\it two} parameters $(s,c_1)$; this will be important for Thomas precession.

\subsection{One-particle states}

Having recalled the classification of BMS$_3$ particles, we now describe their Hilbert space and the transformation law of their wavefunctions. We consider a particle with mass $M$, whose rest-frame supermomentum is $p=M-c_2/24$. Its orbit $\cO_p$ is diffeomorphic to $\Diff/S^1$ and every point $q\in\cO_p$ is obtained by acting on $p$ with a suitable superrotation $g_q$; we refer to the $g_q$'s as {\it standard boosts} and choose them such that they depend smoothly on $q$.\footnote{The manifold $\Diff/S^1$ is homotopic to a point, so globally well-defined standard boosts exist. As for our concrete choice of standard boosts, we will display it in section \ref{sec32}.} The Hilbert space of one-particle states consists of complex square-integrable wavefunctions $\Psi$ on $\cO_p$ that transform under BMS$_3$ as \cite{Wigner:1939cj}
\be
\big(\cU[(f,\alpha)]\cdot\Psi\big)(q)
=
e^{i\langle q,\alpha\rangle}
\cS\big[g_q^{-1}fg_{f^{-1}\cdot q}\big]\Psi(f^{-1}\cdot q).
\label{s12}
\ee
Here $\cU[(f,\alpha)]$ is a unitary operator\footnote{The definition of scalar products and unitarity relies on a measure on $\Diff/S^1$. For simplicity we assume that this measure is superrotation-invariant, though this can be relaxed \cite[chap.\ 3]{Oblak:2016eij}.\label{foot7}} that represents the group element $(f,\alpha)$ and $\cS$ is an irreducible unitary representation of the U$(1)$ group of rotations $\phii\mapsto\phii+\theta$, fixing the particle's spin: $\cS[\theta]=e^{is\theta}$ for some $s\in\RR$. The combination $g_q^{-1}fg_{f^{-1}\cdot q}$ belongs to the little group of $p$ and is known as a {\it Wigner rotation}.\\

For future use it will be important to work out the Lie algebra representation that corresponds to (\ref{s12}) by differentiation. This computation is described in \cite[sec.\ 3.1]{Oblak:2017oge}, to which we refer for details. The flow of any vector field $X$ on $S^1$ defines a one-parameter group of superrotations $e^{tX}$, so we define a representation $\cu$ of the $\bms$ algebra by
\be
\cu[(X,\alpha)]
\equiv
\left.\frac{d}{dt}\right|_0\cU[(e^{tX},t\alpha)].
\label{ss13}
\ee
This is an anti-Hermitian differential operator acting on wavefunctions on $\cO_p$, and its explicit form follows from the terms $e^{i\langle q,\alpha\rangle}$, $\cS[...]$ and $\Psi(f^{-1}\cdot q)$ of eq.\ (\ref{s12}):
\be
\big(\cu[(X,\alpha)]\cdot\Psi\big)(q)
=
\Big(i\langle q,\alpha\rangle+\cs\big[g_q^{-1}\cdot X-g_q^{-1}\text{d}g_q(\xi_X)_q\big]\Big)\Psi(q)-(\xi_X)_q\cdot\Psi.
\label{s13}
\ee
Here $\cs$ is the Lie algebra representation obtained by differentiating $\cS$, so $\cs[...]$ is an imaginary number proportional to the spin $(s,c_1)$ --- see the text right before eq.\ (\ref{s23}). As in (\ref{ss8}), we write $g_q^{-1}\cdot X$ for the action of a diffeomorphism on a vector field on $S^1$, while $\xi_X$ is the fundamental vector field generating the action of $X$ on $\cO_p$:
\be
(\xi_X)_q
\equiv
\left.\frac{d}{dt}\right|_0\Big(e^{tX}\cdot q\Big).
\nn
\ee
The argument of $\cs$ in (\ref{s13}) is the combination $g_q^{-1}\cdot X-g_q^{-1}\text{d}g_q(\xi_X)_q$, which belongs to the Lie algebra of the little group. It is the differential of a Wigner rotation and it will be crucial for Thomas precession.

\section{Berry phases and Thomas precession for BMS$_3$}
\label{sec3}

In this section we act on the rest-frame state of a dressed particle with time-dependent transformations that trace a path in the BMS$_3$ group. This leads to geometric phases in the time evolution of the particle's wavefunction; for closed paths, these become Berry phases \cite{Berry:1984jv} that we evaluate explicitly and relate to Thomas precession. We also show that they are flat limits of analogous Virasoro Berry phases affecting dressed particles in AdS$_3$ \cite{Oblak:2017ect}. Our approach relies on various technical tools relevant to representations of semi-direct products; they are described in greater detail in \cite{Oblak:2017oge}.

\subsection{Berry phases of one-particle states}
\label{sec31}

Let us start with some generalities. Consider a unitary representation $\cU$ of some Lie symmetry group $G$ (with Lie algebra $\mg$) that contains time translations, and think of group elements as `changes of reference frames'. From that perspective, choosing a frame is tantamount to fixing some value for the parameters of a quantum system; varying these parameters in time corresponds to following a path $f(t)$ in $G$. For sufficiently slow variations, the adiabatic theorem \cite{Born1928,Avron:1998th} ensures that such paths only affect energy eigenvectors by multiplication with a geometric phase factor \cite{Berry:1984jv}.\\

Concretely, let $f(t)$ be a closed path in $G$, $t\in[0,T]$, and let $|\phi\ket$ be an eigenstate of the Hamiltonian in a certain reference frame. We assume that the energy $E$ of $|\phi\ket$ is non-degenerate; this will hold for BMS$_3$ particles thanks to their one-dimensional spin space. If the wavefunction at $t=0$ is $\cU[f(0)]|\phi\ket$ and if the path $f(t)$ is traced slowly enough, then the state vector at $t=T$ is
\be
e^{-iET}\,e^{iB[f]}\,\cU[f(0)]|\phi\ket
\nn
\ee
where the first exponential involves the standard dynamical phase $ET$, while the second contains a {\it Berry phase} \cite[sec.\ 2.2]{Oblak:2017ect}
\be
B[f]
=
\oint\text{d}t\,\bra\phi|
i\cu\big[f(t)^{-1}\dot f(t)\big]
|\phi\ket
=
\oint_f\bra\phi|i\cu[f^{-1}\text{d}f]|\phi\ket.
\label{s16}
\ee
Here the integral over $t$ runs from $0$ to $T$ and $\cu$ is the Lie algebra representation corresponding to $\cU$ by differentiation, as in (\ref{ss13}). The $\mg$-valued one-form
\be
f^{-1}\text{d}f
\equiv
\text{d}(L_{f^{-1}})_f
\label{mc}
\ee
is the (left) Maurer-Cartan form of $G$. (We write $L_f$ for left multiplication by $f\in G$.) To apply eq.\ (\ref{s16}) to BMS$_3$ particles, one must first understand how it works for generic semi-direct product groups \cite{Oblak:2017oge}.\\

Accordingly, let $G\ltimes A$ be a semi-direct product, with $A$ a vector group containing time translations. Consider an irreducible unitary representation of $G\ltimes A$ with (super)momentum orbit $\cO_p$, where we think of $p$ as the momentum of a particle in the rest frame. For simplicity we assume that the little group is Abelian so that the spin space is one-dimensional and wavefunctions are valued in $\CC$. In general, the energy spectrum of such a representation is continuous, but the adiabatic theorem still holds \cite{Avron:1998th} provided the variation of the Hamiltonian is sufficiently slow. A related subtlety is that energy eigenstates do not belong to the Hilbert space, as they are not square-integrable; the latter problem can be circumvented by considering smeared wavefunctions \cite{Wu}, so we will think of plane waves with definite momentum as limits of normalized Gaussian wavefunctions. Thus we write $\Phi_k$ for the wavefunction of a particle with (super)momentum $k\in\cO_p$ and normalize it as $\bra\Phi_k|\Phi_k\ket=1$.\footnote{As a wavefunction on $\cO_p$, $\Phi_k(q)=\delta_k(q)/\sqrt{\delta_k(k)}$ where $\delta_k$ is the Dirac distribution at $k$ associated with the measure of footnote \ref{foot7}. The infrared-divergent factor $\delta_k(k)$ enforces the normalization.} In particular, $\Phi_p$ is the wavefunction of a particle at rest. Now, to apply eq.\ (\ref{s16}) we need the Maurer-Cartan form of a semi-direct product such as BMS$_3$. Since the group operation is given by (\ref{ss8}), the Maurer-Cartan form at $(f,\alpha)$ is
\be
(f,\alpha)^{-1}\text{d}(f,\alpha)
=
\big(f^{-1}\text{d}f,f^{-1}\cdot\big)
\nn
\ee
where the two entries on the right-hand side act separately on $T_fG$ and $T_{\alpha}A\cong A$, respectively. Using this together with eq.\ (\ref{s13}), the Berry phase (\ref{s16}) picked up by the wavefunction $\cU\big[\big(f(0),\alpha(0)\big)\big]\Phi_p$ when $(f(t),\alpha(t))$ traces a closed path is
\be
B[f,\alpha]
=
\oint\text{d}t\,\Big<\Phi_p\Big|
\Big(
-\langle p,f^{-1}\cdot\dot\alpha\rangle
+i\cs\big[g_p^{-1}\cdot f^{-1}\dot f-g_p^{-1}\text{d}g_p(\xi_{f^{-1}\dot f})_p\big]-i(\xi_{f^{-1}\dot f})_p
\Big)
\Phi_p\Big>.
\nn
\ee
This expression holds for any group $G\ltimes A$ (provided $g_p^{-1}\cdot f^{-1}\dot f$ is understood as $\Ad_{g_p^{-1}}f^{-1}\dot f$). To simplify it, note that $\bra\Phi_p|(\xi_{\cdots})_p\Phi_p\ket$ is the expectation value of a boost generator in a state at rest, so it vanishes. Using $\bra\Phi_p|\Phi_p\ket=1$, the other two expectation values yield
\be
B_{p,\cs}[f,\alpha]
=
-\oint\text{d}t
\Big(
\langle f\cdot p,\dot\alpha\rangle-i\cs\big[g_p^{-1}\cdot f^{-1}\dot f-g_p^{-1}\text{d}g_p(\xi_{f^{-1}\dot f})_p\big]
\Big)
\label{s17}
\ee
where we write $\langle f\cdot p,\dot\alpha\rangle\equiv\langle p,f^{-1}\cdot\dot\alpha\rangle$. This is the Berry phase formula for semi-direct products \cite[sec.\ 4.2]{Oblak:2017oge}; its spin-dependent term is responsible for Thomas precession.\\

Before we proceed, note that the spin-dependent piece of eq.\ (\ref{s17}) naively seems to depend on standard boosts. Indeed, the corresponding Berry connection is
\be
A_f(\dot f)
=
\cs\big[\Ad_{g_p^{-1}}f^{-1}\dot f-g_p^{-1}\text{d}g_p(\xi_{f^{-1}\dot f})_p\big],
\label{s9b}
\ee
which explicitly involves $g_p$. (We write $g_p^{-1}\cdot X=\Ad_{g_p^{-1}}X$ for any $X\in\mg$.) Since the choice of standard boosts is arbitrary, one might worry that different choices give different Berry phases (\ref{s17}); we now prove that this is not the case --- \ie that (\ref{s17}) is invariant under changes of standard boosts --- assuming for simplicity that the little group is Abelian. Given the $g_q$'s, any other choice of standard boosts takes the form $\tilde g_q=g_qh_q$ where the $h_q$'s belong to the little group of $p$ and depend smoothly on $q$. Using $\tilde g_p^{-1}\text{d}\tilde g_p=\Ad_{h_p^{-1}}\circ(g_p^{-1}\text{d}g_p)+h_p^{-1}\text{d}h_p$, the connection (\ref{s9b}) associated with $\tilde g_q$'s is
\be
\tilde A_f(\cdot f)
=
\cs\Big[\Ad_{h_p^{-1}}\Big(\Ad_{g_p^{-1}}f^{-1}\dot f-g_p^{-1}\text{d}g_p(\xi_{f^{-1}\dot f})_p\Big)-h_p^{-1}\text{d}h_p(\xi_{f^{-1}\dot f})_p\Big].
\nn
\ee
The first term in the argument of $\cs$ reproduces the original expression (\ref{s9b}) up to the adjoint action of $h_p^{-1}$. Since the little group is Abelian, this action is trivial and we get
\be
\tilde A_f(\dot f)
=
A_f(\dot f)
-\cs\big[h_p^{-1}\text{d}h_p(\xi_{f^{-1}\dot f})_p\big].
\nn
\ee
Here the second term is essentially the derivative along $\xi_{f^{-1}\dot f}$ of the scalar function $q\mapsto\log\cS[h_q]$ at $q=p$; it is a total derivative, so its integral over $t$ vanishes when $f(t)$ is periodic. As a result, the Berry phase (\ref{s17}) given by $\tilde A$ is the same as that given by (\ref{s9b}), which proves that (\ref{s17}) is invariant under changes of standard boosts. This will allow us to obtain a general formula for Thomas precession of BMS$_3$ particles, even though its derivation will rely on specific standard boosts.

\subsection{Thomas precession for BMS$_3$ particles}
\label{sec32}

We now apply eq.\ (\ref{s17}) to massive BMS$_3$ particles, taking care to include all central charges. To do this we first need to describe in greater detail the centrally extended BMS$_3$ group (\ref{ss7}). Its elements are quadruples $(f,\lambda,\alpha,\mu)$ where $f$ is a superrotation and $\alpha$ a supertranslation, while $\lambda$ and $\mu$ are real numbers respectively dual to the central charges $c_1$, $c_2$ of eq.\ (\ref{t7}). The group operation extends (\ref{ss8}) and is explicitly given by \cite{Barnich:2014kra}
\be
(f,\lambda,\alpha,\mu)\cdot(g,\rho,\beta,\nu)
=
\Big(f\circ g,\lambda+\rho+C[f,g],\alpha+f\cdot\beta,\mu+\nu-\oint\frac{\text{d}\phii}{24\pi}\,S[f](\phii)\beta(\phii)\Big)
\label{s18}
\ee
where $S[f]$ is the Schwarzian derivative (written below eq.\ (\ref{s10})) while
\be
C[f,g]
=
-\frac{1}{48\pi}\oint\text{d}\phii\log\big(f'(g(\phii))\big)\frac{g''(\phii)}{g'(\phii)}
\label{bocco}
\ee
is the Bott(-Thurston) cocycle \cite{bott1977characteristic} that defines the Virasoro group. In eq.\ (\ref{s18}), the Schwarzian derivative is ultimately responsible for the central charge $c_2$ of (\ref{t7}), while the Bott cocycle yields the superrotational central charge $c_1$.\\

In order to compute Berry phases for BMS$_3$ particles, we separately apply the two terms of (\ref{s17}) to a massive representation. As we shall see, the first, spin-independent part will be sensitive to the massive central charge $c_2$, while the spin-dependent piece will involve $c_1$ and will generalize Thomas precession.

\paragraph{Scalar Berry phase.} In the centrally extended BMS$_3$ group (\ref{ss7}), supertranslations are pairs $(\alpha,\mu)$ as in (\ref{s18}). As a result, their duals are pairs $(p,c_2)$ where $p(\phii)$ is a supermomentum function while $c_2$ is the dimensionful central charge of (\ref{t7}). The pairing is
\be
\big<(p,c_2),(\alpha,\mu)\big>
=
\langle p,\alpha\rangle+c_2\mu
\nn
\ee
where $\langle p,\alpha\rangle$ is the centreless expression (\ref{s9}). To describe Berry phases, we consider closed paths in the group manifold; without loss of generality we take them of the form $(f(t),0,\alpha(t),0)$. Setting $p=M-c_2/24$ to describe a particle with mass $M$, the spin-independent piece of (\ref{s17}) becomes
\begin{align}
\label{pres19}
B_{\text{scalar}}[f,\alpha]
&\stackrel{\text{\textcolor{white}{(10)}}}{=}
-\oint\text{d}t\,\big<f(t)\cdot (p,c_2),(\dot\alpha(t),0)\big>\\
\label{s19}
&\refeq{s10}
-\frac{1}{2\pi}\oint\text{d}t\oint\frac{\text{d}\phii}{f'(t,\phii)}\Big[M-\frac{c_2}{24}+\frac{c_2}{12}S[f(t)](\phii)\Big]\dot\alpha(t,f(t,\phii))
\end{align}
where prime and dot respectively denote partial differentiation with respect to $\phii$ and $t$. This can be split as the sum of a centreless piece, proportional to the mass $M$, and a contribution proportional to the central charge $c_2$:
\be
\boxed{
B_{\text{scalar}}[f,\alpha]
=
-\frac{M}{2\pi}\int\!\text{d}t\,\text{d}\phii\,\frac{\dot\alpha\circ f}{f'}
-\frac{c_2}{24\pi}\int\!\text{d}t\,\text{d}\phii\,\frac{\dot\alpha\circ f}{f'}
\Big(S[f]-\frac{1}{2}\Big)}
\label{ss19}
\ee
where both integrands are evaluated at $(t,\phii)$ and we write $\oint\text{d}t\oint\text{d}\phii\equiv\int\text{d}t\,\text{d}\phii$ for the integral over the torus spanned by $t\in[0,T]$ and $\phii\in[0,2\pi]$. The first term, proportional to $M$, is a standard Berry phase for massive particles, albeit extended to super-rotations and -translations. The second term, proportional to $c_2$, is a genuine gravitational contribution; it vanishes when $(f(t),\alpha(t))$ lies in the Poincar\'e group, since the Schwarzian derivative of any Lorentz transformation (\ref{ss6}) is $S[f]=\tfrac{1}{2}[1-(f')^2]$ so that the time-periodicity of $\alpha(t)$ sets the integral to zero. Note that, thanks to this central term, the Berry phase (\ref{ss19}) is generally non-zero even when $M=0$, in which case it is entirely due to boundary gravitons. It is a flat space analogue of similar phases in AdS$_3$ \cite{Oblak:2017ect} and it coincides with the scalar piece of the BMS$_3$ geometric action derived in \cite{Barnich:2017jgw}. Accordingly, it can also be seen as a symplectic flux on the space of boundary gravitons \cite{boya2001berry,Oblak:2017oge}.

\paragraph{Thomas precession.} The second piece of the Berry phase (\ref{s17}) depends on spin. Its key ingredient is the connection (\ref{s9b}); since the phase is independent of the choice of standard boosts, we assume for simplicity that they reduce to the identity at $p$ so that $g_p=e$. Then the argument of (\ref{s9b}) is a projection of the Maurer-Cartan form (\ref{mc}) of $G$,
\be
f^{-1}\dot f-\text{d}g_p(\xi_{f^{-1}\dot f})_p.
\label{s20}
\ee
In the case at hand we need to evaluate the Maurer-Cartan form of the Virasoro group, whose multiplication is given by the first two entries on the right-hand side of (\ref{s18}). We refer to \cite{Alekseev:1988ce} or \cite[sec.\ 4.3]{Oblak:2017ect} for a derivation. Given a path $f(t,\phii)$ in $\Diff$, the result is
\be
f^{-1}\dot f
\equiv
\left.\frac{\der}{\der\tau}\right|_{\tau=t}
\big(f(t)^{-1},0\big)\cdot\big(f(\tau),0\big)
=
\bigg(\frac{\dot f}{f'}\der_{\phii},
\frac{1}{48\pi}\oint\text{d}\phii\,\frac{\dot f}{f'}\Big(\frac{f''}{f'}\Big)'\,\bigg)
\label{s21}
\ee
where the central piece is a remnant of the cocycle (\ref{bocco}). This is, as it should be, a central extension of the vector field $(\dot f/f')\der_{\phii}$; it is an element of the Virasoro algebra $\hVect$.\\

Now, to evaluate (\ref{s20}) we need to project (\ref{s21}) on the algebra of the little group of a massive particle. The little group consists of rigid rotations, so (\ref{s20}) must be a constant vector field; its form depends on our choice of standard boosts, but this choice is ultimately irrelevant for Berry phases. In practice, to choose these boosts, note that every point $q(\phii)$ on the supermomentum orbit $\cO_p\cong\Diff/S^1$ is uniquely specified by its non-zero Fourier modes, since its energy is then determined by (\ref{ss10}). Thus a natural family of standard boosts is provided by exponentials of vector fields with vanishing average:\footnote{The exponential map from $\Vect$ to $\Diff$ is somewhat pathological \cite[sec.\ 4.4.2]{guieu2007algebre}, so there may be technical subtleties hidden in this choice. We will not address these issues here.}
\be
g_q\in\Diff\text{ std.\ boost}
\quad\Leftrightarrow\quad
g_q=e^{X_q}\text{ with }X_q\in\Vect,\text{ }\oint_{S^1}X_q=0.
\label{ss21}
\ee
In terms of the superrotation generators of eq.\ (\ref{t7}), this means that standard boosts are exponentials of (sums of) $J_n$'s with non-zero $n$, which automatically ensures that $g_p=e$ is the identity. Furthermore, it implies that the second term of (\ref{s20}) is itself a vector field with vanishing average. Indeed, any vector field $X(\phii)\der_{\phii}$ can be split as $X=\tilde X+X_0$ where $X_0$ is its zero-mode, and
\be
\text{d}g_p(\xi_X)_p
=
\left.\frac{d}{dt}\right|_0g_{e^{t(\tilde X+X_0)}\cdot p}
=
\left.\frac{d}{dt}\right|_0g_{e^{t\tilde X+\cO(t^2)}\,\cdot\,e^{tX_0}\,\cdot\,p}
\label{s22}
\ee
where we have used the first term of the Baker-Campbell-Hausdorff formula to split the exponential. Since $X_0$ is constant, it generates rigid rotations and its exponential leaves $p$ invariant; the derivative (\ref{s22}) thus becomes
\be
\text{d}g_p(\xi_X)_p
=
\left.\frac{d}{dt}\right|_0g_{e^{t\tilde X}\,\cdot\,p}
\refeq{ss21}
\left.\frac{d}{dt}\right|_0e^{t\tilde X}
=
\tilde X
=
X-X_0.
\nn
\ee
Accordingly, for exponential standard boosts (\ref{ss21}), the projection (\ref{s20}) of the Maurer-Cartan form coincides with its zero-mode:
\be
f^{-1}\dot f-\text{d}g_p(\xi_{f^{-1}\dot f})_p
\refeq{s21}
\bigg(\frac{1}{2\pi}\oint\text{d}\phii\,\frac{\dot f}{f'}
\;,\;
\frac{1}{48\pi}\oint\text{d}\phii\,\frac{\dot f}{f'}\Big(\frac{f''}{f'}\Big)'\,\bigg).
\label{maca}
\ee
This expression is the argument of the spin representation $\cs$ in the general formula (\ref{s17}). In the present case, $\cs$ is specified by a pair of real numbers $(s,c_1)$ where $s$ is the actual spin of the particle while $c_1$ is the dimensionless central charge of (\ref{t7}). Thus, for any element $(X_0,\lambda)$ of the Lie algebra of the centrally extended little group U$(1)\times\RR$, we have $\cs[(X_0,\lambda)]=i(s-c_1/24)X_0+ic_1\lambda$. Applying this to the projected Maurer-Cartan form (\ref{maca}) and plugging the result in (\ref{s17}), one finds
\be
\boxed{B_{\text{spin}}[f]
=
-\frac{1}{2\pi}\int\!\text{d}t\,\text{d}\phii\,\frac{\dot f}{f'}\bigg[s-\frac{c_1}{24}+\frac{c_1}{24}\Big(\frac{f''}{f'}\Big)'\bigg].}
\label{s23}
\ee
This is the Berry phase due to the spin of a massive BMS$_3$ particle. It comes from the spin-dependent term of (\ref{s17}), which in turn originates from the Wigner rotation $g_q^{-1}fg_{f^{-1}\cdot q}$ of eq.\ (\ref{s12}). In this sense, the phase is the angle of a finite rotation that results from the composition of a sequence of infinitesimal Wigner rotations; it is a generalization of the usual phenomenon of Thomas precession \cite{Thomas:1926dy}.\\

\begin{figure}[t]
\centering
\includegraphics[width=0.30\textwidth]{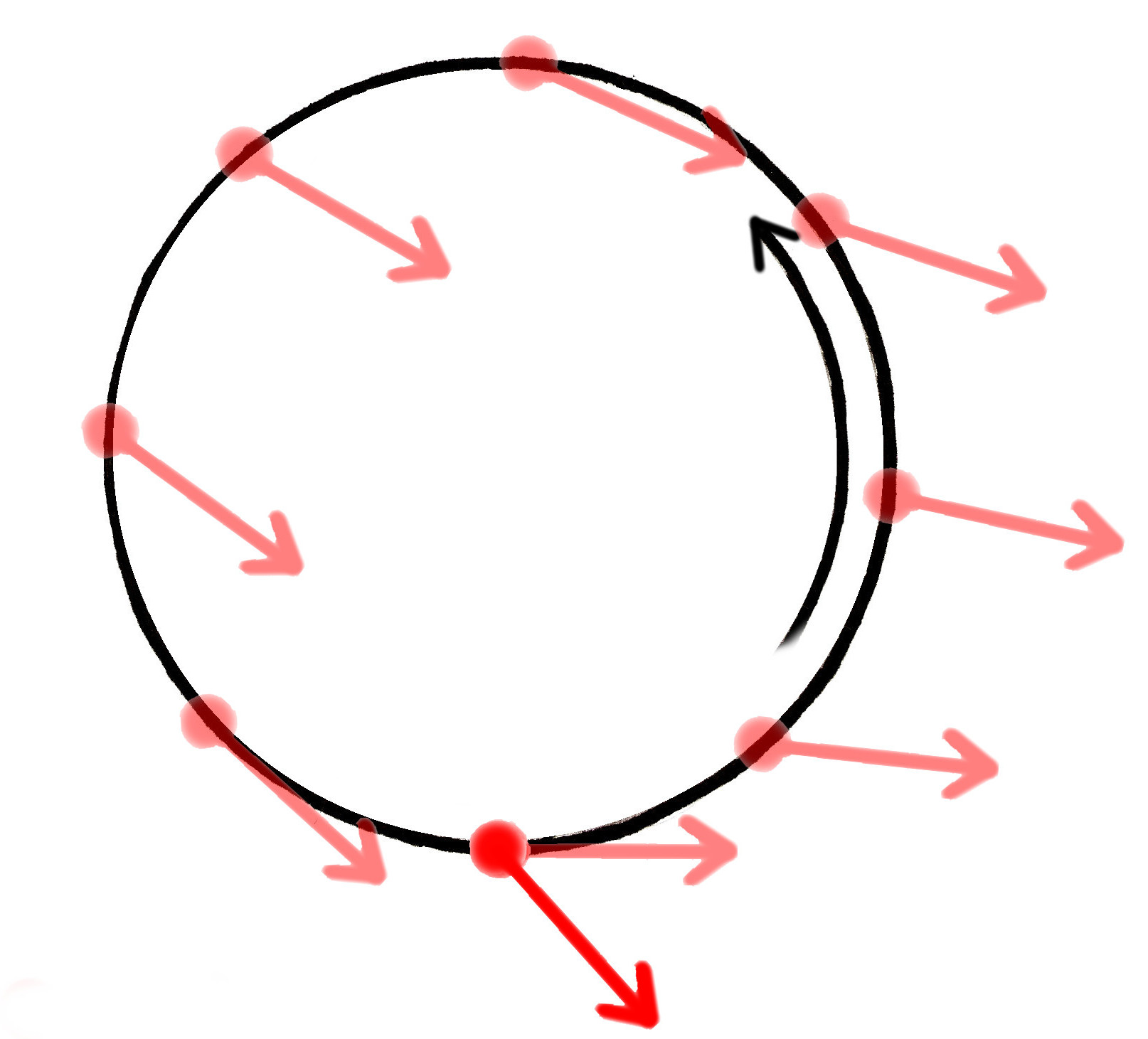}
\caption{A particle with spin $s>0$ follows a (counterclockwise) circular trajectory at constant rapidity. The red arrow represents the phase of its wavefunction. This phase Thomas-precesses in a clockwise direction; after one revolution, it has turned by an angle (\ref{bini}). The general formula (\ref{tt24}) extends this phenomenon to superrotations.\label{t24}}
\end{figure}

Note that (\ref{s23}) vanishes when $f(t,\phii)=\phii+\theta(t)$ consists only of rotations. This is a special case of a more general statement: for any periodic time-dependent rotation $h(t)$, the paths $f(t)$ and $f(t)\circ h(t)$ give the same Berry phase. As a result, the functional (\ref{s23}) is best seen as a function on the quotient space $\Diff/S^1$ rather than $\Diff$ itself. The former is the orbit of a massive BMS$_3$ particle, so one can think of of (\ref{s23}) as a phase shift that occurs when a particle follows a closed path $f(t)\cdot p$ in supermomentum space (see fig.\ \ref{t24}). In particular one can even define Berry phases for {\it open} paths $f(t)$ in $\Diff$, as long as $f(0)^{-1}\circ f(T)$ is a pure rotation --- what matters is only that the {\it projection} of $f(t)$ on $\Diff/S^1$ be closed. This subtlety is explained in greater detail in \cite[sec.\ 2.2]{Oblak:2017ect}.\\

To confirm the relation with Thomas precession, let us apply eq.\ (\ref{s23}) to specific loops in the group of superrotations. For instance, suppose that $f(t,\phii)$ is a rotating path
\be
f(t,\phii)
=
g(\phii)+\omega t
\label{s24}
\ee
with $\omega>0$ an angular velocity, and $g(\phii)$ a `superboost' of order $n\in\NN$ and rapidity $\lambda$:
\be
e^{ing(\phii)}
=
\frac{\cosh(\lambda/2)e^{in\phii}+\sinh(\lambda/2)}{\sinh(\lambda/2)e^{in\phii}+\cosh(\lambda/2)}.
\label{ss24}
\ee
The curve (\ref{s24}) is such that $f(0)^{-1}\circ f(t)$ becomes a pure rotation at $t=2\pi/\omega$. At that point, the path $f(t)\cdot p$ closes in supermomentum space and the Berry phase is \cite{Oblak:2017ect}
\be
B_{\text{spin}}[f]
=
-2\pi\Big(s+\frac{c_1}{24}(n^2-1)\Big)(\cosh\lambda-1).
\label{tt24}
\ee
For $n=1$, the transformation (\ref{ss24}) is a Lorentz boost (see eq.\ (\ref{ss6})) and the contribution of the central charge vanishes:
\be
B_{\text{Thomas}}
=
-2\pi s\,(\cosh\lambda-1).
\label{bini}
\ee
This Berry phase is a rewriting of the standard formula for Thomas precession of a particle with spin $s$ following a circular trajectory at rapidity $\lambda$. It says that the particle's wavefunction undergoes a net, non-zero rotation when the particle completes one revolution in momentum space. The minus sign accounts for the fact that a trajectory traced in a counterclockwise direction results in a {\it clockwise} precession; see fig.\ \ref{t24}.\\

Formula (\ref{s23}) has already appeared in the literature, starting with \cite{Alekseev:1988ce}, where it was seen as an action functional for a field $f(t,\phii)$ associated with a Virasoro representation at central charge $c_1$ and highest weight $s$. A similar viewpoint was recently put forward in \cite{Barnich:2017jgw}, where (\ref{s23}) is the spin-dependent piece of an action for boundary gravitons in three dimensions. For our purposes however, the closest interpretation is that of \cite{Oblak:2017ect}, where a similar Berry phase was derived for dressed particles in AdS$_3$. Indeed the results (\ref{s23})-(\ref{tt24}) are identical to their AdS$_3$ peers (eqs.\ (43)-(48) of \cite{Oblak:2017ect}) up to the replacement of $(h,c)$ by $(s,c_1)$. On group-theoretic grounds, this matching was expected: since superrotations span a Virasoro group, the resulting Berry phases must closely resemble their pure Virasoro version. This being said, we stress that the derivation of BMS$_3$ Berry phases described here follows a very different logic than that of the Virasoro case; in particular, the latter involves no standard boosts or Wigner rotations, hence no projection (\ref{s20}).

\subsection{Flat limit of Virasoro Berry phases}

It was shown in \cite{Barnich:2012aw} that the phase space of three-dimensional flat space gravity is a limit of its AdS$_3$ counterpart as the cosmological constant goes to zero. In the same vein, the Berry phases (\ref{ss19})-(\ref{s23}) should follow from a large $c$ limit of their Virasoro peers \cite{Oblak:2017ect}. The purpose of this section is to verify this explicitly. Spoiler alert: the limit will work perfectly, so the hasty reader may skip the next few pages and go straight to section \ref{sec4}.\\

We consider gravity in AdS$_3$ with a cosmological constant $\Lambda=-1/\ell^2$. For suitable fall-off conditions on the metric \cite{Brown:1986nw}, the asymptotic symmetries of the system span two copies of the Virasoro group\footnote{Other fall-offs may lead to enhanced symmetries \cite{Troessaert:2013fma}, but they always contain two Virasoro's.} that act on light-cone coordinates $x^{\pm}$ on the cylinder at infinity (the boundary of AdS$_3$) as conformal transformations $x^+\mapsto F(x^+)$, $x^-\mapsto\bar F(x^-)$. Their action on phase space is projective, with left and right central charges $c$, $\bar c$. In Einstein gravity one has $c=\bar c=3\ell/2G$, but if parity is broken one can have $c\neq \bar c$ \cite{Hotta:2008yq}. Now, a particle dressed with boundary gravitons in AdS$_3$ is a unitary representation of the direct sum of two Virasoro algebras with highest weights $h,\bar h$; its spin is $s=h-\bar h$ and its mass is $(h+\bar h)/\ell$ up to quantum corrections. When acted upon by a cyclic family of conformal transformations $(F(t,x^+),\bar F(t,x^-))$, its state vector picks up a Berry phase \cite{Oblak:2017ect}
\be
B
=
-\int\frac{\text{d}t\,\text{d}x^+}{2\pi}
\frac{\dot F}{\der F}\left[
h-\frac{c}{24}
+
\frac{c}{24}\,\der\!\left(\frac{\der^2F}{\der F}\right)\,
\right]
-\int\frac{\text{d}t\,\text{d}x^-}{2\pi}
\frac{\dot{\bar F}}{\bar\der\bar F}\left[
\bar h-\frac{\bar c}{24}
+
\frac{\bar c}{24}\,\bar\der\!\left(\frac{\bar\der^2\bar F}{\bar\der\bar F}\right)\,
\right]
\label{s27}
\ee
where $\der\equiv\der/\der x^+$, $\bar\der\equiv\der/\der x^-$. The function $F$ satisfies $\der F>0$ and $F(x^++2\pi)=F(x^+)+2\pi$, and the same holds for $\bar F$ in terms of $x^-$. The integrals over $t$ run from $0$ to $T$ and those over $x^{\pm}$ run from $0$ to $2\pi$.\\

We now ask whether formula (\ref{s27}) has a well-defined flat space limit. This limit is achieved when the cosmological constant goes to zero so that $\ell/G$ goes to infinity; for brevity we simply write this as $\ell\rightarrow\infty$. We define
\be
c_1\equiv\lim_{\ell\rightarrow\infty}(c-\bar c),
\qquad
c_2\equiv\lim_{\ell\rightarrow\infty}\frac{c+\bar c}{\ell}
\label{circa}
\ee
and require that $c,\bar c$ scale with $\ell$ in such a way that both of these quantities be finite. We also require the particle's spin and mass to be finite in the limit,
\be
s\equiv\lim_{\ell\rightarrow\infty}(h-\bar h),
\qquad
M\equiv\lim_{\ell\rightarrow\infty}\frac{h+\bar h}{\ell}.
\label{seh}
\ee
Finally, following \cite{Barnich:2012aw} we think of the AdS$_3$ light-cone coordinates as $x^{\pm}=\frac{u}{\ell}\pm\phii$ where $(u,\phii)$ will eventually be the flat space Bondi coordinates of eq.\ (\ref{tt6}). The only missing piece is the relation between the conformal transformation $(F,\bar F)$ and the BMS$_3$ transformation $(f,\alpha)$; to describe it we let $F,\bar F$ depend parametrically on $\ell$ and expand them as
\be
\begin{split}
F(x^+) &= F_0(\phii)+\frac{u}{\ell}F_0'(\phii)+\frac{1}{\ell}F_1(\phii)+\cO(1/\ell^2),\\[.1cm]
\bar F(x^-) &= \bar F_0(\phii)-\frac{u}{\ell}\bar F'_0(\phii)+\frac{1}{\ell}\bar F_1(\phii)+\cO(1/\ell^2).
\end{split}
\label{expa}
\ee
Here the terms with derivatives come from the Taylor expansion of $F_0(x^+)$ and $\bar F_0(x^-)$, while $F_1,\bar F_1$ account for the parametric expansion of $F,\bar F$ in powers of $1/\ell$. In order to relate these unknown functions to a BMS$_3$ group element $(f,\alpha)$, we require that the map $x^+\mapsto F(x^+)$, $x^-\mapsto\bar F(x^-)$ reproduces (\ref{t6}) in the large $\ell$ limit. The result is
\be
F_0(\phii)
=
-\bar F_0(\phii)
=
f(\phii),
\qquad
\demi\big(F_1(\phii)+\bar F_1(\phii)\big)=\alpha(f(\phii)).
\nn
\ee
Note that this does not fix the difference $F_1-\bar F_1$, which is arbitrary. It turns out that this difference does not affect Berry phases at all, so we set it to zero without loss of generality. In consequence the expansions (\ref{expa}) can be written as
\be
F(x)
=
f(x)+\frac{\alpha(f(x))}{\ell},
\qquad
\bar F(x)
=
-f(-x)+\frac{\alpha(f(-x))}{\ell}
\label{s14b}
\ee
where we neglect terms of order $1/\ell^2$. In both equations, $x$ is an arbitrary real argument, $\alpha$ is $2\pi$-periodic, while $f(x)$ has a positive derivative and satisfies $f(x+2\pi)=f(x)+2\pi$. Furthermore, $f$ and $\alpha$ both depend on the time parameter $t$ in a $T$-periodic way.\\

It now remains to plug (\ref{s14b}) in the Berry phase (\ref{s27}), write the result in terms of flat space data, and take the large $\ell$ limit. For convenience we write $f(-x)\equiv \bar f(x)$ and denote derivatives with respect to $x$ by a prime. In addition we rename the integration variables $x^{\pm}$ of (\ref{s27}) into $x$. Then $\der F$ and $\bar\der\bar F$ are replaced by $F'$ and $\bar F'$, and the quantities appearing in the centreless piece of (\ref{s27}) are
\be
\frac{\dot F}{F'}
=
\frac{\dot f}{f'}+\frac{1}{\ell}\,\frac{\dot\alpha\circ f}{f'},
\qquad
\frac{\dot{\bar F}}{\bar F'}
=
\frac{\dot{\bar f}}{\bar f'}-\frac{1}{\ell}\,\frac{\dot\alpha\circ\bar f}{\bar f'}.
\nn
\ee
As a result, the centreless part of (\ref{s27}) has a well-defined large $\ell$ limit:
\be
-\int\frac{\text{d}t\,\text{d}x}{2\pi}\bigg[\frac{\dot F}{F'}(h-c/24)+\frac{\dot{\bar F}}{\bar F'}(\bar h-\bar c/24)\bigg]
\stackrel{\ell\rightarrow\infty}{\longrightarrow}
-\int\frac{\text{d}t\,\text{d}x}{2\pi}\bigg[\frac{\dot\alpha\circ f}{f'}(M-c_2/24)+\frac{\dot f}{f'}(s-c_1/24)\bigg]
\nn
\ee
where we used (\ref{circa}) and (\ref{seh}) to replace $(c,\bar c,h,\bar h)$ by $(c_1,c_2,s,M)$. This limit is the sum of the centreless parts of the flat space Berry phases (\ref{s19}) and (\ref{s23}). A similar method can be applied to the central terms of (\ref{s27}), involving third derivatives of $F,\bar F$. For example, the leading terms in (\ref{s14b}) yield $F=f$ and $\bar F=-\bar f$ up to $1/\ell$ corrections, which produces the central part of the spinning Berry phase (\ref{s23}):
\be
-\frac{1}{2\pi}\int\!\text{d}t\,\text{d}x
\bigg[\frac{c}{24}\frac{\dot f}{f'}\bigg(\frac{f''}{f'}\bigg)'+\frac{\bar c}{24}\frac{\dot{\bar f}}{\bar f'}\bigg(\frac{\bar f''}{\bar f'}\bigg)'\bigg]
\stackrel{\ell\rightarrow\infty}{\longrightarrow}
-\frac{1}{2\pi}\int\!\text{d}t\,\text{d}\phii\,\frac{c_1}{24}\frac{\dot f}{f'}\bigg(\frac{f''}{f'}\bigg)'.
\nn
\ee
Finally, the central piece of the scalar Berry phase (\ref{s19}) is obtained by including the $1/\ell$ part of (\ref{s14b}) in the central terms of (\ref{s27}); the computation is conceptually identical to those above, if mildly more technical. The end result is that the large $\ell$ limit of the AdS$_3$ Berry phase (\ref{s27}) coincides with the sum of (\ref{ss19}) and (\ref{s23}), which was to be proved.

\section{Discussion}
\label{sec4}

In this paper we have shown how cyclic sequences of BMS$_3$ transformations generate Berry phases in the wavefunction of a dressed particle. According to the general formula (\ref{s17}), each such phase is the sum of two terms:
\be
B[f,\alpha]
=
B_{\text{scalar}}[f,\alpha]+B_{\text{spin}}[f].
\label{bespin}
\ee
\begin{itemize}
\item The scalar contribution is displayed in (\ref{ss19}) and involves both superrotations and supertranslations. It is blind to spin but it is sensitive to the mass $M$ and the dimensionful central charge $c_2$. The latter is essentially the Planck mass (\ref{s8}), so the Berry phase is generally non-zero even when $M=0$; in that case it is a contribution of gravitational dressing to the vacuum wavefunction.
\item The spin-dependent piece (\ref{s23}) involves only superrotations, and is a gravitational extension of Thomas precession. It is only sensitive to the spin $s$ and the dimensionless central charge $c_1$. The latter vanishes in Einstein gravity, in which case
\be
B_{\text{spin}}[f]=-\frac{s}{2\pi}\int\!\text{d}t\,\text{d}\phii\,\frac{\dot f}{f'}
\qquad\text{(for $c_1=0$)}.
\label{biless}
\ee
When $f(t,\phii)$ consists of Lorentz transformations (\ref{ss6}), this reduces to known formulas for Thomas precession. Note that the phase (\ref{biless}) is also non-zero for paths that consist of genuine superrotations (which are not in the Lorentz group).
\end{itemize}
These results are consistent with their AdS$_3$ analogues \cite{Oblak:2017ect}, of which they are a limit, although the tools \cite{Oblak:2017oge} needed for BMS$_3$ differ sharply from their Virasoro counterparts. Note also that (\ref{bespin}) coincides with the kinetic action functional recently derived in \cite{Barnich:2017jgw}, so that gravitational Berry phases can be seen as symplectic fluxes on the space of boundary gravitons. In the remainder of this section we discuss some prospects for actually observing such Berry phases and briefly describe their four-dimensional version.

\subsection{Are these phases measurable?}

Our motivation for investigating Berry phases for Virasoro and BMS was that they potentially provide measurable quantities associated with asymptotic symmetries. The obvious problem then is to understand how such phases could be observed (if at all). Unfortunately we will not provide a definite, realistic answer to this question, but let us at least try to devise a {\it gedankenexperiment} to guide our intuition. Thomas precession normally occurs when a spinning particle such as an electron orbits around an atomic nucleus (fig. \ref{t24}); this contributes to its effective spin-orbit coupling, which in turn affects spectral lines (see \eg \cite[sec.\ 11.8]{Jackson:1998nia}). Now, the revolution of the particle around the nucleus can be interpreted as a sequence of Lorentz boosts, so the gravitational generalization of Thomas precession could presumably be switched on when the particle is subjected to a sequence of asymptotic symmetry transformations. These transformations, in turn, can be triggered by suitable field excitations crossing null infinity; this is the essence of the memory effect \cite{zel1974radiation} and of its relation to BMS symmetry \cite{Strominger:2014pwa}. So in short, gravitational Berry phases should be switched on when a disturbance of the space-time metric propagates in a way that generates time-dependent, cyclic asymptotic symmetry transformations.\\

In practice, the model studied in this paper and its precursor \cite{Oblak:2017ect} is three-dimensional gravity, which has no local degrees of freedom. Accordingly, the only way to generate a memory effect is to couple gravitation to some other field (\eg a scalar or a gauge field), and consider configurations such that the back-reacted space-time metric achieves the desired asymptotic symmetry transformation. In fact, part of the recipe for doing this has already appeared in the literature in the guise of BMS$_3$ fluid dynamics at null infinity \cite{Penna:2017vms}. In the latter reference, the metric undergoes a time-dependent family of BMS$_3$ transformations generated by a flow $\zeta_u$; if it were possible to find bulk field configurations giving rise to a cyclic flow, then the resulting transformations $(f(t),\alpha(t))$ should lead to Berry phases (or, classically, to Hannay angles). Another, possibly more realistic prospect, is to measure symmetry-based Berry phases in two-dimensional CFT; this might be feasible in the quantum Hall effect, where a Berry connection very similar to the one that gives (\ref{s23}) has appeared in \cite{Bradlyn:2015wsa}.

\subsection{Berry phases in four dimensions}

Since the focus of this paper was BMS symmetry, it is natural to ask whether our considerations extend to realistic four-dimensional space-times, where gravitational waves actually exist \cite{Abbott:2016blz} so that gravitational Berry phases might occur without the need for extra fields in the bulk. Technically speaking, this is a much harder problem than in three dimensions because the representations of BMS$_4$ are much less understood than those of BMS$_3$. In particular, the unitary representations built by McCarthy \cite{McCarthy01} do not appear to account for gravitational dressing, which is crucial for soft theorems \cite{Strominger:2013jfa}, and include neither superrotations \cite{Barnich:2009se} nor a central charge \cite{Barnich:2011mi,Barnich:2017ubf}. Nevertheless, the structure of eq.\ (\ref{s19}) above is simple enough to guess how its centreless part generalizes to BMS$_4$. We now briefly describe this generalization.\\

Just as BMS particles in three dimensions, their counterparts in four dimensions should be classified by orbits of supermomenta (Bondi mass aspects) under either superrotations or Lorentz transformations. The quantum states of any such particle are wavefunctions on a supermomentum orbit, whose transformation law under supertranslations should be (see eq.\ (\ref{s12}))
\be
\Psi(q)\mapsto e^{i\langle q,\alpha\rangle}\Psi(q)
\nn
\ee
where $\langle q,\alpha\rangle$ is the surface charge associated with the supertranslation $\alpha$ when the Bondi mass aspect is the function $q$ on the celestial sphere. Since BMS$_4$ is a semi-direct product, the arguments of section \ref{sec31} should apply to its representations and lead, in particular, to a scalar Berry phase of the form (\ref{pres19}). In the case at hand the circle $S^1$ gets replaced by a sphere $S^2$ with stereographic coordinate $z\in\CC$, so that
\be
B_{\text{scalar}}[f,\alpha]
=
\frac{1}{2\pi i}\oint\text{d}t\int_{S^2}\frac{dzd\bar z}{(1+z\bar z)^2}\big(f(t)\cdot p\big)(z,\bar z)\,\dot\alpha\big(t,f(z),\bar f(\bar z)\big).
\label{basca}
\ee
Here $f\cdot p$ denotes the action of a (time-dependent) superrotation $f$ on the Bondi mass aspect $p(z,\bar z)$ \cite{Barnich:2016lyg}, but one can also restrict attention to the Lorentz subgroup where
\be
f(t,z)=\frac{a(t)z+b(t)}{c(t)z+d(t))}
\nn
\ee
is a global conformal transformation under which $p(z,\bar z)$ is a quasi-primary field with weights $(3/2,3/2)$. With this restriction, the central extension of BMS$_4$ does not appear but eq.\ (\ref{basca}) still generalizes standard Berry phases in a non-trivial way thanks to the (time-dependent) supertranslation $\alpha$. Thus, as in three dimensions, dressed particles in four dimensions should generally pick up Berry phases when acted upon by a cyclic family of BMS supertranslations accompanied by superrotations or Lorentz boosts.\\

When it comes to superrotations, the four-dimensional case is much more subtle than its three-dimensional cousin. Indeed, in contrast to BMS$_3$, the action of four-dimensional superrotations on celestial spheres is singular \cite{Barnich:2011mi}; this is in fact the same situation as in Euclidean two-dimensional CFTs, except that here the singularities appear at null infinity and have much more dramatic effects \cite{Strominger:2016wns}. Therefore, it is not clear if a formula such as (\ref{s23}) has any chance to work in the four-dimensional realm; in particular, Thomas precession may not admit any extension due to superrotations. In addition, even without dealing with superrotations, an extra subtlety of BMS$_4$ symmetry is the presence of a field-dependent central extension \cite{Barnich:2011mi,Barnich:2017ubf} that should play a key role for representation theory. Regardless of these puzzles, our main goal in this work was to point out that gravitational dressing can, in principle, lead to observable Berry phases. This conclusion should remain true in four dimensions, at least qualitatively; hopefully, time will tell whether more quantitative arguments can be found to support this proposal.

\newpage
\section*{Acknowledgements}

I am grateful to A.~Campoleoni, M.~Godazgar and A.~Seraj for inspiring discussions and to T.~Neupert and F.~Schindler for clarifications regarding the quantum Hall effect. I also wish to thank G.~Mullier for sharing with me his Batman symbol. This work is supported by the Swiss National Science Foundation, and also partly by the NCCR SwissMAP.

\addcontentsline{toc}{section}{References}

\end{document}